\documentclass{ecai}

\usepackage{latexsym}
\usepackage{amssymb}
\usepackage{amsmath}
\usepackage{amsthm}
\usepackage{booktabs}
\usepackage{enumitem}
\usepackage{graphicx}
\usepackage{color}
\usepackage{adjustbox}

\usepackage{times}
\usepackage{soul}
\usepackage{url}
\usepackage[hidelinks]{hyperref}
\usepackage[utf8]{inputenc}
\usepackage[small]{caption}
\usepackage{algorithmic}

\newtheorem{example}{Example}

\newcommand{\BibTeX}{B\kern-.05em{\sc i\kern-.025em b}\kern-.08em\TeX}

\newcommand{\miko}{Mikol\'a\v s Janota}
\newcommand{\joao}{Jo\~ao  Ara\'ujo}
\newcommand{\ccw}{Choiwah Chow}
\usepackage{mathtools}
\usepackage{graphicx}

\usepackage{booktabs}
\usepackage{threeparttable}
\usepackage[linesnumbered,ruled,vlined]{algorithm2e}
\DontPrintSemicolon%
\SetKwProg{Func}{Function}{}{}%
\SetKwComment{tcc}{/*}{*/}%
\SetKwInOut{Input}{input}\SetKwInOut{Output}{output}%
\SetKwData{false}{false}\SetKwData{true}{true}
\SetKwData{done}{done}
\SetKwData{newCubes}{newCubes}
\SetKwData{nonIsoCube}{nonIsoCube}
\SetKwData{C}{C}
\SetKwData{D}{D}
\SetKwData{E}{E}
\SetKwData{H}{H}
\SetKwData{models}{models}
\SetKwData{void}{void}
\SetKwFunction{Canonisize}{Canonicalize}
\SetKwFunction{Canonicalize}{Canonicalize}
\SetKwFunction{NextCell}{NextCell}
\SetKwFunction{Search}{Search}

\newcommand{\seqnum}[1]{\href{https://oeis.org/#1}{\rm \underline{#1}}}

\newcommand{\cbif}{\texttt{CBIF}\xspace}
\newcommand{\mace}{\texttt{Mace4}\xspace}
\newcommand{\nauty}{\texttt{nauty}\xspace}
\newcommand{\Nauty}{\texttt{Nauty}\xspace}

\newcommand{\loops}{\texttt{LOOPS}\xspace}
\newcommand{\isonaut}{\texttt{Isonaut}\xspace}
\newcommand{\isofilter}{\texttt{Isofilter}\xspace}
\newcommand{\isofiltertwo}{\texttt{Isofilter2}\xspace}
\newcommand{\valtrue}{\texttt{True}\xspace}
\newcommand{\valfalse}{\texttt{False}\xspace}

\newcommand{\tpl}[1]{\ensuremath{\langle #1 \rangle}}

\newtheorem{notation}{Notation}

\makeatletter
\newcommand*\bigcdot{\mathpalette\bigcdot@{.5}}
\newcommand*\bigcdot@[2]{\mathbin{\vcenter{\hbox{\scalebox{#2}{$\m@th#1\bullet$}}}}}
\makeatother

\usepackage{tikz}
\usepackage{pdfpages}
\usetikzlibrary {matrix}
\usetikzlibrary {arrows.meta}
\usetikzlibrary{calc, patterns}
\usetikzlibrary{shapes.misc}
\tikzset{cross/.style={cross out, draw=black, minimum size=2*(#1-\pgflinewidth), inner sep=0pt, outer sep=0pt},	
	cross/.default={5pt}}
\tikzstyle{finmod} = [%
font={},
column 1/.style={font=\bfseries},
row 1/.style={font=\bfseries},
inner sep=0pt, 
matrix of nodes,draw=none,nodes={%
	text width=1.5em,align=center, minimum height=1.0em,anchor=center}
]
\usepackage{pgfplots}
\pgfplotsset{compat=1.15}

\definecolor{lightgreen}{rgb}{.9,1,.9}
\definecolor{darkblue}{rgb}{0,0,0.6}
\definecolor{darkred}{rgb}{0.7,0,0}

\begin{document}


\begin{frontmatter}


\paperid{627}


\title{Cube-based Isomorph-free Finite Model Finding}


\author[A]{\fnms{Choiwah}~\snm{Chow}\orcid{0000-0002-2067-0568}\thanks{\ccw. Email: choiwah.chow@gmail.com}} 
\author[A]{\fnms{Mikol\'a\v s}~\snm{Janota}\orcid{0000-0003-3487-784X}\thanks{\miko. Email: mikolas.janota@gmail.com.}} 
\author[B]{\fnms{Jo\~ao}~\snm{Ara\'ujo}\orcid{0000-0001-6655-2172}\thanks{\joao. Email: jjrsga@gmail.com}}

\address[A]{Czech Technical University in Prague, Czech Republic}
\address[B]{Universidade Nova de Lisboa, Lisbon, Portugal}


\begin{abstract}
	Complete enumeration of finite models of first-order logic (FOL) formulas is pivotal
to universal algebra, which studies and catalogs algebraic structures.
Efficient finite model enumeration is highly challenging because the number of models 
grows rapidly with their size but at the same time, we are only interested in models
modulo isomorphism. While isomorphism cuts down the number of models of interest,
it is nontrivial to take that into account computationally.

This paper develops a novel algorithm that achieves
isomorphism-free enumeration by employing isomorphic graph detection algorithm
nauty, cube-based search space splitting, and compact model representations. We
name our algorithm cube-based isomorph-free finite model finding algorithm
(CBIF). Our approach contrasts with the traditional two-step algorithms, which
first enumerate (possibly isomorphic) models and then filter the isomorphic ones out
in the second stage.
The experimental results show that CBIF is many orders of magnitude faster than
the traditional two-step algorithms. CBIF enables us to calculate new results that are
not found in the literature, including the extension of two existing OEIS sequences, thereby advancing the state of the art.

\end{abstract}

\end{frontmatter}


\section{Introduction}%
Enumerating finite algebras is a very important task in universal algebra. E.g., after observing that there is only one group of prime order, it is easy to conjecture, and then prove, that there is only one group of order (size) $p$, for prime $p$.  Model enumeration has mostly been performed by
the two-step direct-search model enumeration procedure, which is plagued by isomorphic models.
Intuitively, two models are isomorphic if one model can be transformed to the other by renaming its domain elements.
More formally, two models are isomorphic if one can be transformed to the other by an isomorphism, which is a structure preserving bijective function.
That is, isomorphism is an equivalence relation that partitions the set of models into equivalence classes.  We need only one member from each of these equivalence classes, and the rest are noise that need to be removed. In our example of group of prime order, if isomorphic models are allowed in model enumeration, there will be more than one group for each order.  We may miss the hint that there is only one group of any prime order.
Furthermore, in many instances, it is crucial for mathematicians to possess not just the number, but the complete list of all non-isomorphic models of a certain size $n$ within a given theory. The guarantee that no models have been omitted makes these libraries invaluable, as it enables mathematicians to draw definitive conclusions from testing conjectures on them. Consequently, significant effort is devoted to creating and maintaining these comprehensive libraries.
Libraries such as \texttt{SmallGrp} \cite{smallgrp}, \texttt{Smallsemi} \cite{smallsemi0.6.12}, and \texttt{Loops} \cite{loops3.4.1} provide comprehensive lists of groups, semigroups, and other algebraic structures, ensuring that mathematicians have access to complete datasets. All these libraries have been produced with a combination of deep theorems and dedicated computational tools, which are not always available. Thus, mathematicians need simple and efficient tools that are not dedicated to just a few structures to produce libraries of all non-isomorphic models of a given theory for any given size.
However, isomorphic models are often generated in large proportions during the first step of the non-dedicated model search process, and it is difficult to identify which pairs of models are isomorphic.
For example, the traditional model generator Mace4 generates over 423 million models for the Involutive Lattices of order 13, belonging to only 8,028 isomorphism classes.  That is, less than 0.002\% of the models generated are non-isomorphic to each other. It takes the computer in our experiments (see Section~\ref{sec:experiments}) 2.5 days to generate these $423$+ million models according to the definition of Involutive Lattices (the first step) and many more hours to filter out the isomorphic models (the second step).
\emph{In this paper we propose to generate the non-isomorphic models
  directly without the intermediate outputs, which take time to generate,
  require disk space to hold, and more time to filter out the redundant
isomorphic models.}

If the model search process can be organized as a search tree, then
a common strategy is to prune this search tree as much as possible and as early as possible during the search process. We
can prune a branch of the search tree if we know that it does not produce any model that is not isomorphic to those already found in other searched branches.
If we prune all the isomorphic branches during the search process, then we have an isomorph-free model search algorithm. We will introduce in Section~\ref{sec:algorithm} the \emph{cube-based isomorph-free finite model search algorithm} (\cbif), which efficiently identifies and removes isomorphic branches and models during the search process.

Fast isomorph-free algorithms are paramount in
computational universal algebra.
However, as previouly pointed out, few algebras have specialized algorithms for generating models. \cbif, on the other hand, provides mathematicians with a generic tool that enables enumerating models beyond the state of the art.

The main contributions of this paper are the following:
\begin{enumerate}
  \item We introduce a novel approach that applies the \nauty algorithm~\cite{mckay2022nauty} in conjunction with a data encoding scheme and a hash table for fast detection of isomorphic cubes (Section~\ref{sec:prelim:subsec:FME}) and models.
  \item Based on the new isomorphic cubes/model detection algorithm, we devise a novel cube-based isomorph-free finite model searching algorithm, \cbif.
  \item We provide an implementation of the algorithm, which enables us to calculate new results that are not yet reported in the literature .
\end{enumerate}

\section{Preliminaries}%
\label{sec:prelim}
We will be concerned only with models definable in first-order logic (FOL) with equation.
We follow well-established terminology, cf.~\cite{araujo_et_al:LIPIcs.CP.2023.8,02661029-c12f-3b75-a049-79273fea242e,DBLP:conf/birthday/Zhang013}, in this paper. Models are defined by a signature $\Sigma$ and a FOL formula $\mathcal{F}$ on $\Sigma$, which may contain $k$-ary function symbols, and $m$-ary predicate (or relations) symbols, where $k, m\in\mathbb{N}_0$. Nullary functions are called \emph{constants}. Predicates are treated as functions with the values \valtrue and \valfalse.
So our description on functions applies to predicates as well.
The domain is denoted by the set $D=\{0, \dots, {n}-1\}$, where $n\ge 2$. The size of $D$ is called the \emph{order} of the model.

\subsection{Finite Model Enumeration}%
\label{sec:prelim:subsec:FME}

A traditional direct-search finite model finder first expands the FOL formula to its ground representation by its domain elements in $D$. Then it searches for models by assigning values to the functions and predicates in $\mathcal{F}$, backtracking to the previous step upon failure, or when all values are exhausted, and retrying with different values.

A \emph{value assignment (VA) clause} is a term $f(a_1, \dots, a_k)=v$, where~$f$ is a $k$-ary function symbol in $\Sigma$ and $a_j, v\in D$. The term
$f(a_1, \dots, a_k)$ is called a \emph{cell term} (or simply \emph{cell}).
Conceptually, the finite model finder fills the cells of the operation table of~$f$ with values one by one.

To search for models in $\mathcal{F}$, the direct-search finite model finder employs a \emph{function selection} strategy in conjunction with a \emph{cell selection} strategy to pick cell terms successively, without duplicates, to assign values from $D$ to form VA clauses.
\begin{example} \label{example:cell:selection}
	Suppose we are searching for models of order $n$ with just one binary operation $f$. Then a possible cell selection strategy is to order the cells by their first function arguments, then by their second function arguments. So, the cells will be selected in the following order: $f(0,0), f(0,1), \cdots, f(0, n-1), f(1, 0), \cdots, f(n-1, n-1)$.	The traditional finite model finder assigns values from $D$ to the cells in the list successively.   If all cells are assigned values without violating the axioms set forth in $\mathcal{F}$, then a model is found. \qed
\end{example}

The cell selection strategy has big impacts on the performance of the finite model finite, but it is beyond the scope of this paper to discuss it in details.

After a value is assigned, a finite model finder may optionally do propagation.  There are two kinds of propagation. Positive propagation allows the inference of values for some empty cells.  Negative propagation excludes the possibility of some values being assigned to some empty cells. Both kinds of propagation speed up the search process.

\begin{example}
	(Positive Propagation) Suppose the FOL formula contains the equation $f(x,y) = f(y,x)$, that is, the operation $f$ is commutative.  After the assignment $f(a,b)=c$, where $a, b, c\in D$, the finite model finder can infer $f(b,a)=c$ and make the assignment immediately.     \qed
\end{example}

\begin{example}
	(Negative Propagation) Suppose the FOL formula contains the inequality $f(x,y)\neq f(y,x)$, then after the assignment $f(a,b)=c$, where $a, b, c\in D$, the finite model finder does not need to try to assign $c$ to the cell $f(b,a)$.  If all but one domain elements are eliminated by negative propagation for a cell, then the finite model finder can assign the only remaining value to that cell.   \qed
\end{example}

After the optional propagation step, the axioms in $\mathcal{F}$ are checked against this new VA clause and any other propagated VA clauses. If any of the axioms are violated, then the VA clause, along with any propagated assignments, are reverted. A new value will be tried for that cell and the process continues. If no value can be assigned to that cell term without failing the axioms in $\mathcal{F}$, then the model finder backtracks to the previous cell to try to assign another value to it.  When all cell terms in $\mathcal{F}$ are assigned values and the axioms hold, a model represented by its VA clauses is found. After that, the process may continue with backtracking to find more models.

The models produced this way often contain a lot of redundancies. If a model can be transformed to another model by renaming its domain elements, then one of them is redundant because it offers no new information about the underlying structure. Formally, renaming domain elements is achieved by applying an isomorphism (a structure-preserving bijective function) to the models. Two models are said to be isomorphic to each other if an isomorphism exists from one model to the other.
Note that the permutation cycle $(a, b)$, denoted by $\pi_{(a,b)}$, can be used as a function for renaming $a$ to $b$ and vice versa. For example, all three models in Figure~\ref{fig:example:iso} are isomorphic. Model $A$ can be transformed to model $B$ by $\pi_{(2,3)}$,
and model $B$ can be transformed to model $C$ by $\pi_{(1,3)}$. $C$ turns out to be the lexicographically smallest model in the equivalence class containing these three models, and is usually the preferred model to keep~\cite{DBLP:conf/aaai/JanotaC0CV24}.\\

\begin{figure}[htb]
	\centering
	\caption{Isomorphic models $A$, $B$, and $C$}\label{fig:example:iso}
	\begin{tikzpicture}[outer sep=0, inner sep=5]
		\matrix [finmod, label={[font=\small]above:Model $A$}] ($A$) at (0, 0) {
			$*_A$ & 0  & 1 & 2 & 3  \\
			0  & 0 & 1 & 0 & 3 \\
			1  & 1 & 2 & 1 & 2 \\
			2  & 2 & 1 & 2 & 1  \\
			3  & 3 & 0 & 3 & 0  \\
		};

		\matrix [finmod, label={[font=\small]above:Model $B$}] ($B$) at (3, 0) {
			$*_B$ & 0 & 1 & 2 & 3  \\
			0  & 0 & 1 & 2 & 0 \\
			1  & 1 & 3 & 3 & 1 \\
			2  & 2 & 0 & 0 & 2  \\
			3  & 3 & 1 & 1 & 3  \\
		};

		\matrix [finmod, label={[font=\small]above:Model $C$}] ($C$) at (6, 0) {
		$*_C$ & 0 & 1 & 2 & 3  \\
		0  & 0 & 0 & 2 & 3 \\
		1  & 1 & 1 & 3 & 3 \\
		2  & 2 & 2 & 0 & 0  \\
		3  & 3 & 3 & 1 & 1  \\
		};

		\foreach \name in {$A$,$B$,$C$} {%
			\draw[gray] (\name-5-1.south east) to (\name-1-1.north east); 
			\draw[gray] (\name-1-1.south west) to (\name-1-5.south east); 
		}
	\end{tikzpicture}
\end{figure}
\

Another way to look at the direct-search model finding process is through a search tree. The search space can be organized as a search tree in which nodes are VA clauses and edges join successive nodes with cell terms in the search order. The root node is an empty VA clause.  A \emph{search path} in a search tree is a path from the root to a node in the search tree. It can be represented by a sequence of VA clauses $\tpl{t_0=v_0; t_1=v_1; \cdots}$, where $t_i$ is the cell term in the $i^{th}$ position of the sequence and $v_i\in D$. Furthermore, $t_i\neq t_j$ when $i\ne j$. A search path is terminated at the first VA clause that results in a violation of any axiom of $\mathcal{F}$.

A useful notation here is the \emph{cube}, which is defined as the prefix of a search path.
Cubes facilitate massive parallelization (e.g.,~\cite{araujo_et_al:LIPIcs.CP.2023.8}) although parallelization is not attempted in this version of \cbif that follows the recorded object paradigm.
Being a prefix of a search path, a cube can be specified by a sequence of VA clauses.
Conceptually, the finite model finder extends the cube one VA clause at a time until a model is found. Isomorphisms, and permutations, can be applied to a cube by applying them to its VA clauses. Specifically, if $\pi$ is a permutation on $D$, and $B$ is a cube, then 
	$\pi(B) \coloneqq \{f(\pi(a_1), \dots, \pi(a_{k}))= \pi(v)\ |\ f(a_1, \dots, a_{k})=v \text{ is a VA clause in B} \}$.

We define $\pi(b)=b$ for $b\in\{\valtrue,\valfalse\}$ for any isomorphism $\pi$ since \valtrue and \valfalse need to be preserved.
We use $\pi_{id}$ to denote the identity permutation. Observe that $\pi_{id}(B)$ is the set of all individual VA clauses in the cube $B$.

Cubes are said to be isomorphic to each other if their VA clauses are isomorphic to each other. In particular, two cubes $B_0$ and $B_1$ are isomorphic if there is a permutation $\pi$ on $D$ such that $\pi(B_0)=\pi_{id}(B_1)$. %
\begin{example}\label{ex:longer_cubes}
	If $B_0=\tpl{f(0)=0; g(0,0)=0; f(1)=0}$ and $B_1=\tpl{f(0)=1; g(0,0)=1; f(1)=1}$, then $B_0$ and $B_1$ are not isomorphic because no isomorphism can be found to transform $B_0$ to $B_1$. However,
	extending each cube with one more VA clause s.t.\ $B_0=\tpl{f(0)=0; g(0,0)=0; f(1)=0; g(1,1)=0}$ and $B_1=\tpl{f(0)=1; g(0,0)=1; f(1)=1; g(1,1)=1}$, then $B_0$ and $B_1$ are isomorphic because $\pi_{(0,1)}(B_0)=\{f(1)=1, g(1,1)=1, f(0)=1, g(0,0)=1\}=\pi_{id}(B_1)$.   \qed
\end{example}

\section{Graph Isomorphism}%
\label{sec:iso:naut}
A cornerstone of our \cbif algorithm is the fact that isomorphic cubes extend to isomorphic models~\cite[Theorem~9]{araujo_et_al:LIPIcs.CP.2023.8}.
Thus, a cube can be excluded from the search if it is isomorphic to another cube already covered. We incorporate the graph isomorphism checking tool \nauty~\cite{mckay2022nauty} into \cbif to speed up the isomorphic cubes/models detection process. Since \nauty works on graphs,
we need to first construct a graph corresponding to the cube/model before \nauty can be applied. 
Graph construction algorithm for models are only briefly, and often incompletely, described in the literature, e.g.,~\cite{DBLP:journals/access/Khan20a,automatedmodel2015}. So, we give a complete description of our version of the procedure, including how relations are handled, before we describe how to extend it to also handle cubes.


Suppose the model $M$ is defined by a signature $\Sigma$ and a FOL formula $\mathcal{F}$ on $\Sigma$. Suppose further that $\mathcal{F}$ consists of a collection of functions $f_i$ of arity $k_i$, and relations $r_j$ of arity $m_j$, where $i$, $j$, $k_i$, and $m_j$ are non-negative integers.  Let $q$ be the highest arity of functions and relations in $\mathcal{F}$. $G$ then comprises a collection of sets of vertices and a collection of edges between some of these vertices. Furthermore, each set of vertices is of a different color.  \Nauty does not rename vertices with different colors to each other so as to prevent the renaming of, for example, vertices representing cell terms to vertices representing function values. First, some notations: 

\begin{notation}
	Let $X$ be a set of vertices in $G$. Then $X_d$ denotes the vertex in $X$ associated with the domain element $d$.
\end{notation}

\begin{notation}
	$A_p$ denotes the set of vertices for the $p^{th}$ argument of the functions and relations in $\mathcal{F}$.
\end{notation}

Thus, ${A_p}_d$ denotes the vertex in $A_p$ corresponding to the domain element $d$.

\begin{notation}
	Suppose $f_i$ is a function or relation in $\mathcal{F}$ and $f_i(a_1,\cdots,a_{k_i})$ is a cell, then $V_{f_i}$ denotes the set of vertices for the $i^{th}$ function or relation (in any fixed order) in $\mathcal{F}$, and $V_{f_i(a_1,\cdots,a_{k_i})}$ denotes the vertex of the cell  $f_i(a_1,\cdots,a_{k_i})$.
\end{notation}

\noindent
The vertices of $G$ are constructed as follow:
\begin{enumerate}
	\item For each domain element $d$ add a vertex $E_d$ to the set of vertices $E$ of $G$. 
	\item For each domain element $d$ and each $p$, where $1\le p \le q$,
	add a vertex ${A_p}_d$ to the set of vertices $A_p$. 
	\item For each domain element $d$, add a vertex $R_d$ to the set of vertices $R$.   These vertices represents function values. 
	\item  Add to $G$ two vertices $B_T$ and $B_F$, each with a distinct color from the rest of the graph.
	\item For each cell term, $f_i(a_1, \cdots, a_{k_i})$,
	add a vertex $V_{f_i(a_1,\cdots,a_{k_i})}$ to the set of vertices $V_{f_i}$. 
	\item For each cell term, $r_j(a_1,\cdots,a_{m_j})$,
	add a vertex $V_{r_j(a_1,\cdots,a_{m_j})}$ to the set of vertices $V_{r_j}$. 
\end{enumerate}

The purpose of the edges is to connect directly or indirectly all vertices corresponding to the same domain element, so that if a domain element is renamed by an isomorphism, then all its vertices are renamed accordingly because of the connections.
So, we add edges to $G$ the following way:
\begin{enumerate}
	\item Add an edge between each pair of vertices $E_d$ and ${A_p}_d$, where $1 \le d \le n$ and $1 \le p \le q$.
	\item Add an edge between each pair of vertices $E_d$ and $R_d$, where $1 \le d \le n$.
	\item For functions:
	For each VA clause, $f_i(a_1,\cdots,a_{k_i})=d$:
	\begin{enumerate}
	   \item add an edge between $V_{f_i(a_1,\cdots,a_{k_i})}$ and $R_d$.
	   \item add an edge between each pair of the vertices $V_{f_i(a_1,\cdots,a_{k_i})}$ and ${A_p}_d$, where $1 \le d \le n$.
	\end{enumerate}
	\item For VA clause $r_j(a_1,\cdots,a_{m_j})=B$ on predicate $r_j$: 
	\begin{enumerate}
		\item add an edge between the vertices $V_{r_j(a_1,\cdots,a_{m_j})}$ and $B_T$ (if $B=\valtrue$) or $B_F$ (if $B=\valfalse$).
		\item add an edge between each pair of the vertices $V_{r_j(a_1,\cdots,a_{m_j})}$ and ${A_p}_d$, where $1 \le d \le n$.
	\end{enumerate}
\end{enumerate}



\begin{figure}[!hbt]
    \centering
    \vspace{1em}
	\caption{Graph of Model $M$}\label{fig:graph:m}
	\begin{tikzpicture}[xscale=0.6]
\clip (-1,-1.1) rectangle (12, 5.5);
\draw (4.5, 4) node {$E$};
\node[draw, shape = rectangle, fill=lightgreen] (E0) at (5, 3.5) {$0$};
\node[draw, shape = rectangle, fill=lightgreen] (E1) at (5.78, 3.5) {$1$};
\draw (4.5, 2.5) node {$A_2$};
\node[draw, shape = rectangle, fill=green] (S0) at (5, 2) {$0$};
\node[draw, shape = rectangle, fill=green] (S1) at (5.78, 2) {$1$};
\draw (-0.5, 3) node {$A_1$};
\node[draw, shape = rectangle, fill=lime] (F0) at (0.5, 2.5) {$0$};
\node[draw, shape = rectangle, fill=lime] (F1) at (0.5, 2) {$1$};
\draw[thin] (E0) -- (S0);
\draw[thin] (E0) -- (F0);
\draw[thin] (E1) -- (S1);
\draw[thin] (E1) to[out=110,in=170] (F1);
\draw (10.5, 3) node {$R$};
\node[draw, shape = rectangle, fill=orange] (R0) at (11, 2.5) {$0$};
\node[draw, shape = rectangle, fill=orange] (R1) at (11, 2) {$1$};
\draw[thin] (E0) to[out=100,in=90] (R0);
\draw[thin] (E1) to[out=315,in=170] (R1);
\draw (4.5, 1.05) node {$V_f$};
\node[draw, shape = rectangle, fill=yellow] (V00) at (5, 0.51) {$0$};
\node[draw, shape = rectangle, fill=yellow] (V01) at (5.78, 0.51) {$1$};
\node[draw, shape = rectangle, fill=yellow] (V10) at (5, 0) {$0$};
\node[draw, shape = rectangle, fill=yellow] (V11) at (5.78, 0) {$1$};
\draw[thin] (F0) -- (V00);
\draw[thin] (F1) to[out=0,in=180] (V10);
\draw[thin] (S0) -- (V00);
\draw[thin] (S1) -- (V01);
\draw[thin] (R0) to[out=200,in=50] (V00);
\draw[thin] (R1) to[out=190,in=50] (V01);
\draw[thin] (R1) to[out=230,in=-5] (V11);
\draw[thin] (R0) to[out=-10,in=280] (V10);
\draw[thin] (S0) to[out=180,in=180] (V10);
\draw[thin] (S1) to[out=0,in=0] (V11);
\draw[thin] (F1) to[out=190,in=310] (V11);
\draw[thin] (F0) to[out=-20,in=125] (V01);
\end{tikzpicture}
    \vspace{1em}
\end{figure}
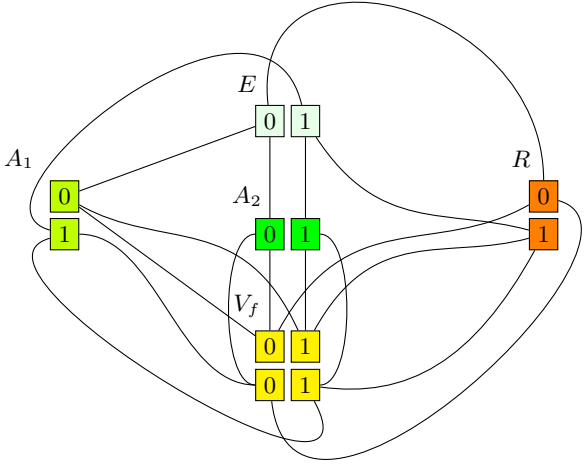

\begin{example}\label{graph:example}
Take the model $M$ of order 2 defined by one binary operation $f$: $\tpl{f(0,0)=0, f(0,1)=1, f(1,0)=0, f(1,1)=1}$. Figure~\ref{table:model:m} shows the operation table for $f$. Its graph $G$ constructed from the procedure described above is illustrated in Figure~\ref{fig:graph:m}. \Nauty uses consecutive non-negative integers (starting from zero) to represent vertices, and edges are represented a pair of vertices.  The vertices of the graph generated by the above procedure are:
$E=\{0, 1\}$, $A_1=\{2, 3\}$, $A_2=\{4, 5\}$, $R=\{6, 7\}$, and $V_f=\{8, 9, 10, 11\}$.  Each set $E$, $A_1$, $A_2$, $R$, and $V_f$ is given a different color.
The edges are represented by their end points, as shown in Figure~\ref{fig:vertices:m}.  There are  12 vertices and 18 edges in the graph. \Nauty prints out an undirected edge $(v_1, v_2)$ only when $v_1<v_2$, and prints $v_1$ only once for all undirected edges connected to it, as illustrated in Figure~\ref{fig:vertices:m}.  For example, the first line on the left side of the figure states that there is an edge between each of the pairs of vertices $0$ and $4$, $0$ and $2$, and $0$ and $6$.

Let's use the VA clause $f(0,1)=1$ to illustrate how the edges are constructed. This VA clause is represented by the yellow square (vertex) on the first row, second column of $V_f$. It has an edge to the vertex marked as ``$0$'' in $A_1$ because the first argument of $f$ is $0$. It has an edge to the vertex marked as ``$1$'' in $A_2$ because the second argument of $f$ is $1$. Finally, it has an edge to the vertex marked as ``$1$'' in $R$ because the value assigned to the cell term is $1$.
\qed
\end{example}

\begin{figure}
\begin{minipage}[!th]{0.15\textwidth}
	\vspace{-1.6cm}
\caption{Operation Table of $M$}
		\begin{tikzpicture}[baseline]
			\matrix [finmod] (B) at (0, 0) {
				$f$ & 0 & 1\\
				0 & 0 & 1\\
				1 & 0 & 1\\
			};
			\foreach \name in {B} {%
			\draw[gray] (\name-3-1.south east) to (\name-1-1.north east); 
			\draw[gray] (\name-1-1.south west) to (\name-1-3.south east); 
			}
		\end{tikzpicture}
		\label{table:model:m}
\end{minipage}
\hspace{0.1cm}
\begin{minipage}[ht]{0.3\textwidth}

\hspace{0.1cm}
\centering
\caption{Vertices of $M$}
\vspace{0.3cm}
\begin{minipage}[ht]{0.3\textwidth}
\vspace{0.3cm}
0 : 4 2 6\\
1 : 5 3 7\\
2 : 8 11\\
3 : 9 10\\
4 : 8 9\\
5 : 10 11 \\
\end{minipage}
\hspace{0.1cm}\vspace{0.3cm}
\vline\hspace{0.4cm}
\begin{minipage}[!ht]{0.3\textwidth}
6 : 8 10\\
7 : 9 11\\
8 :\\
9 :\\
10 :\\
11 :
\end{minipage}
\label{fig:vertices:m}
\end{minipage}
\end{figure}

The steps in constructing the graph $G$ for a cube is exactly the same as those for a model, except that an additional vertex $U$ corresponding to the \emph{unassigned} value is needed because not all cells are assigned values. We also need to add an edge between $U$ and each cell, $f_i(a_1,\ldots,a_{k_i})$ or $r_j(a_1,\ldots,a_{m_j})$, that is not yet assigned a value.  Note that $U$ does not have any edge to vertices in $E$, $R$, $B_T$, $B_F$, or $A_p$ because it is not a domain element or a truth value.

\begin{example}\label{example:sparse:graph}
For the model $M$ in Example~\ref{graph:example}, there are exactly 3 edges coming out of each vertex in its graph $G$ (see Figure~\ref{fig:graph:m}). So there are $3*12/2=18$ edges for the graph of $M$ of $12$ vertices. A complete undirected graph with $12$ vertices has $12(12-1)/2=66$ edges. So $G$ is a sparse graph.  In fact, the graph of the model becomes  sparser as the order goes higher. Thus, \nauty has to be run in the sparse mode for the best performance.
\end{example}



\section{Checking for Isomorphisms}%
\label{sec:checking:isomorphisms}
Given two models or cubes, we can construct their representations in graphs and apply \nauty to them to find their canonical graphs.  Models/Cubes are isomorphic if and only if they have the same canonical graph.  However,
a direct comparison of the canonical graphs requires $n(n-1)/2$ comparisons for $n$ graphs in the worse case scenario. The canonical graphs are also quite big, and needs to stay in the memory for fast comparisons. We use a hashing scheme and a special compact representation of models to mitigate these two pitfalls.

Example~\ref{graph:example} shows that the vertices (Figure~\ref{fig:vertices:m}) representing a model require much more space than the operations tables. In that example, the binary operation table (Figure~\ref{table:model:m}) can be written out as a sequence of 4 numbers, e.g., as an ASCII string ``0101''. The graph, on the other hand, contains 30 numbers.  There are ways to reduce the size of the graph, e.g., by omitting the left-most column of numbers as they are simply an increasing sequence of consecutive integers.  However, it is still much bigger than the 4 numbers in the binary operation table.

\begin{figure}[!htb]
	\caption{Model Dataflow}\label{fig:model:dataflow}
	\begin{tikzpicture}[xscale=0.9,yscale=0.9]
\clip (-1.5,-2.4) rectangle (8, 5.7);

\node[text width=4cm] (M0) at (1, 5) {Model $M_0$ represented by an ordered list of operation tables.};
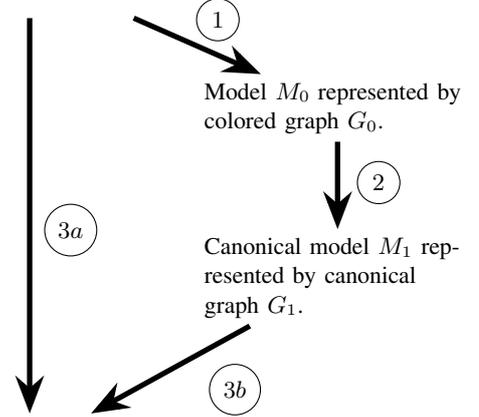
\node[text width=3.5cm] (G0) at (5.5, 3) {Model $M_0$ represented by colored graph $G_0$.};
\draw[line width=2pt, -{Stealth[length=5mm]}] (M0) -- (G0);
\node[draw, shape = circle] (Step1) at (3.75, 4.3) {$1$};
\node[text width=3.5cm, shape=rectangle] (G1) at (5.5, 0.5) {Canonical model $M_1$ represented by canonical graph $G_1$.};
\draw[line width=2pt, -{Stealth[length=5mm]}] (G0) -- (G1);
\node[draw, shape = circle] (Step2) at (6.1, 1.9) {$2$};
\node[text width=4cm] (M1) at (1, -2.0) {Canonical model $M_1$ as an ordered list of operation tables.};
\draw[line width=2pt, -{Stealth[length=5mm]}] (G1) -- (M1);
\draw[line width=2pt, -{Stealth[length=5mm]}] (M0) -- (M1);
\node[draw, shape = circle] (Step3a) at (1.6, 1.2) {$3a$};
\node[draw, shape = circle] (Step3b) at (4, -1.15) {$3b$};
\end{tikzpicture}
\end{figure}

To cut down the memory footprint of \cbif, we work with the model's entities (functions, relations etc.) in identifying isomorphic models and cubes. The steps are illustrated in Figure~\ref{fig:model:dataflow}. Given a model $M_0$, we construct its corresponding graph $G_0$ using the algorithm given in Section~\ref{sec:iso:naut} (Step~1 in Figure~\ref{fig:model:dataflow}). We then apply \nauty to $G_0$ to get its canonical representation $G_1$ (Step 2). \Nauty also outputs the isomorphism $I$ from $G_0$ to $G_1$. Let $M_1$ be the model that gives $G_1$ when the algorithm given in Section~\ref{sec:iso:naut} is applied to it.  So we can apply $I$ to $M_0$ to arrive at the model $M_1$ (the step marked as 3a and 3b).  Note that the graph $G_1$ of model $M_1$ is in its canonical representation (by construction). We call $M_1$ the \emph{canonical model} of $M_0$. All models having the same canonical model are isomorphic to each other. We can therefore use $M_1$ instead of $G_1$ to identify isomorphism in models.

For any  model $M_0$, we use the encoded operation tables of its canonical model $M_1$ as its key in hashing. 
Since cubes are just partial models, so the same algorithm apply in isomorphism of cubes.
Now a cube/model can be encoded as an ordered list of operation tables, and each operation table can be encoded as an ordered list of cell values as fixed-width numbers. The output of this encoding scheme is just one long sequence of numbers. For example, the model in Example~\ref{graph:example} could be represented by the 4-byte string ``0101'', or by any other fixed width sequence of numeric values such as a bit vector. This compact encoding scheme greatly reduces the footprint of the memory used for isomorphic models/cubes detection, which is important in recorded object algorithms.

\begin{example}
Consider the 2 models $D$ and $E$ as shown in Figure~\ref{fig:example:canon}. Suppose the graphs of $D$ and $E$ are $G_D$ and $G_E$, respectively.

\begin{figure}[htb]
	\centering
	\caption{Models $D$ and $E$}\label{fig:example:canon}
	\begin{tikzpicture}[outer sep=0, inner sep=5]
		\matrix [finmod, label={[font=\small]above:Model $D$}] ($D$) at (0, 0) {
			$*_D$ & 0  & 1  \\
			0  & 0 & 0 \\
			1  & 0 & 1 \\
		};

		\matrix [finmod, label={[font=\small]above:Model $E$}] ($E$) at (3, 0) {
			$*_E$ & 0 & 1   \\
			0  & 0 & 1 \\
			1  & 1 & 1 \\
		};

		\foreach \name in {$D$,$E$} {%
			\draw[gray] (\name-3-1.south east) to (\name-1-1.north east); 
			\draw[gray] (\name-1-1.south west) to (\name-1-3.south east); 
	}
\end{tikzpicture}
\end{figure}

When \nauty is applied to $G_D$, we obtain the canonical graph $G$, and also the identity permutation.  So, the canonical model of $D$ is $D$ itself. When \nauty is applied to the graph $G_E$, we obtain the same canonical graph $G$, and the permutation $\pi_{(0, 1)}$.  Applying $\pi_{(0, 1)}$ to the model $E$ gives us the model $D$.  That is, the canonical model of both $D$ and $E$ are the same, and we therefore conclude that $D$ and $E$ are isomorphic. This example shows that we can compare canonical models instead of canonical graphs for isomorphism. The advantages of using canonical models are that they are much smaller and faster to compare than their corresponding canonical graphs.

\end{example}


\section{Isomorph-free Model Finding}%
\label{sec:algorithm}

\cbif searches for models by forming VA clauses one by one as in any direct-search finite model finder.  During its traversal of the search tree, it records the canonical cubes and models in a hash table. It discards the cube or model if its canonical version is already in the hash table.  It thus avoids extending any cube that is isomorphic to a cube that has previously been explored.  We know that this shortcut does not lose any models because isomorphic cubes only lead to isomorphic models, which have to be eliminated~\cite{araujo_et_al:LIPIcs.CP.2023.8}.

We summarize this approach in Algorithm~\ref{algorithm:generate:models}, in which the function  \texttt{NextCell} returns the next cell to explore (void if none left), and the function  \texttt{Canonicalize}  returns the canonical model/cube of the cube being explored. For simplicity, it is written as a depth-first search algorithm, but breath-first search or any other traversal order also works. It enumerates all models, but it can be changed to stop after finding one or any specified number of models.

\begin{algorithm}[!htb]
	\caption{Cube-based Isomorph-free Model Generation}%
	\label{algorithm:generate:models}
	\Input{Axioms $\mathcal{F}$ and Domain elements \D}
	\Output{A set of non-isomorphic models}
	$\H \gets $ empty hash table \;
	
	\Func{$\Search(\C, t)$} {%
		\tcc{\C is a cube, $t$ is a cell term}
		$\models \gets \emptyset$ \;
		\ForEach{$v \in \D $}  {  %
			$\E \gets \C \cup \{t=v\}$ \;
			\If{\E does not violate $\mathcal{F}$ and $\Canonicalize(\E) \notin \H$}{%
				$\H \gets \H \cup\Canonicalize(\E)$ \;
				$t \gets\NextCell(\E)$ \;
				\If{$t$ is $\void$} {%
					$\models \gets \models \cup \E$ \;
				}
				\Else{%
					$\models\gets\models\cup\Search(\E, t)$ \;
				}
			}
		}
		\Return \models \;
	}
	
	
	\Return $\Search(\emptyset, \NextCell(\emptyset))$ \;
\end{algorithm}

Example~\ref{ex:longer_cubes} in Section~\ref{sec:prelim:subsec:FME} illustrates an important property of isomorphic cubes: more isomorphic cubes can be found in longer cubes. This means the \cbif retains its efficacy when applied to models of higher orders.




\section{Experimental Results}%
\label{sec:experiments}
We incorporate \cbif into the finite model enumerator \mace, which supports searching on FOL~\cite{Mccune_2003_tech264}. 
It is considered one of the best finite model finders. 
The current prototype~\cite{mace4c:source} supports operations up to arity of 2 but it
can be extended to support operations of any arities. For comparison, a separate standalone program, \isonaut~\cite{isonaut:source}, is developed to use the same procedure as \cbif to remove isomorphic models from a set of models. In our implementation of the two-step algorithm, we use \mace to generate models according to the given FOL, and then use \isonaut to remove isomorphic models from the outputs of \mace. \Nauty is run in the \emph{sparse} mode because the graphs from cubes/models are sparse graphs (see Example~\ref{example:sparse:graph}).
The experiments\footnote{Data and scripts are available in~\cite{experiments:source}.} are run on an Intel\textregistered\ Xeon\textregistered Silver 4110 CPU\@ 2.0 GHz $\times$32 computer, with 64 GB RAM.  All times reported here are CPU times. Except for the IP loops of order 15, we allow 1.5 days for each experiment to run.

The MarcieDB database~\cite{marciedb} contains a rich collection of many popular algebras suitable for our benchmarking purposes. 
We pick a wide range of challenging algebraic structures with different complexities for our experiments. 
We run the tests with higher orders to highlight the power of the \cbif algorithm. %
The definitions of the algebras used in the experiments in this section are listed in Table~\ref{table:algebra:definition}, in which all clauses are implicitly universally quantified.
  
\begin{table}[!htb]
	\caption{Definitions of Algebras Used in Experiments}
	\begin{tabular}{ll}
		\toprule
		\multicolumn{1}{p{1cm}}{Algebra}  & \multicolumn{1}{p{3cm}}{FOL Definition}   \\ \cmidrule{1-2}
		Tarski & $(x * y) * y = (y * x) * x.$\ \ $(x * y) * x = x.$\\
		Algebras &$x * (y * z) = y * (x * z). $ \\ \cmidrule{1-2}
		Involutive & $(x * y) * z = x * (y * z).$\ \ \ $x * y = y * x.$ \\
		Lattices& $(x + y) + z = x + (y + z).$\ \ \ $--x = x.$\\
		& $x + y = y + x.$\ \ \ \ $(x * y) + x = x.$ \\ 
		&$(x + y) * x = x.$\ \ \ $-(x + y) = -x * -y.$ \\ \cmidrule{1-2}
		M-zeroids & $(x + y) + z = x + (y + z).$\\
		&$(x * y) * z = x * (y * z).$\ \ \ $x+y=y+x$. \\
		&$x+0 = x.$\ \ \ $(x \times y) \times z = x + (y \times z).$   \\		
		& $(x * y) \times x = x.$\ \ \ $(x \times y) * x = x.$\\
		&$-x = x.$\ \ \ $x \times y=y \times x$. \ \ \  $x+-x = 0.$\\
		&$x * y = y*x.$\ \ \ $x < y \leftrightarrow 0 = -x + y.$ \\
		& $x + (y * z) = (x + y) * (x + z).$\\ \cmidrule{1-2}
		Near-rings & $(x + y) + z = x + (y + z).$\ \ \ $x + 0 = x.$\\
		&$x + - x = 0$.\ \ \ $(x * y) * z = x * (y * z)$. \\
		&$(x + y) * z = (x * z) + (y * z)$.\\ \cmidrule{1-2}
		Tarski's & $(x + y) + z = x + (y + z).$\ \ $x*1=1*x.$\\
		HSI &$(x * y) * z = x * (y * z)$.\ \ $x * y = y * x$.\\
		&$x * (y + z) = (x * y) + (x * z)$. \ \ $x^1=x.$\\
		&$(x^y) * (x^z) = x^{(y+z)}.$\ \ \ $(x*y)^z = x^z * y^z.$ \\ 
		&$(x^y)^z = x^{(y*z)}.$\ \ $1^x=x.$\ \ $x + y = y + x.$ \\ \cmidrule{1-2}
		Loops & $x * y = x * z \rightarrow y = z.$\ \ $0 * x = x.$ \\
		& $y * x = z * x \rightarrow y = z.$\ \ $x * 0 = x.$ \\  \cmidrule{1-2}
		C-loops & Loops with  $x * (y * (y * z)) = ((x * y) * y) * z.$ \\  \cmidrule{1-2}
		IP loops & Loops with  $x' * (x * y) = y.$\ \ $(y * x) * x' = y.$ \\ 
		& Redundant clauses to speed up search: $0*0=0.$~$0'=0.$ \\ \bottomrule
	\end{tabular}
	\label{table:algebra:definition}
\end{table}


We first run tests on 3 algebras with increasing complexities to show the improvements in performance of the \cbif algorithm over the two-step finite model finding algorithm.

Speedup is calculated as the ratio of the time of the two-step model finding algorithm and the time of the  \cbif algorithm on the same data set.  That is, it is the number of times that \cbif is as fast as the conventional two-step method.

\begin{figure}[!htb]
	\centering
	\caption{Performance of \cbif on Models of Higher Orders}

\begin{tikzpicture}[scale=0.75, every node/.style={scale=0.9}]
        \begin{axis}[
                xlabel={Order},
                ylabel={Time in Seconds (log scale)},
                xmin=6, xmax=24,
                ymin=0.1, ymax=250000,
                xtick={6,10,15,20,24},
                ytick={0.1,1,10,100,1000,10000,100000},
                legend pos=north west,
                ymajorgrids=true,
                ymode=log,
                grid style=dashed,
                ]

                \addplot[
                color=black,
                mark=triangle,
                ]
                coordinates {
                        (9,0.28)(10,1.61)(11,2.5)(12,16.6)(13,28.77)(14,187)(15,359)(16,2368)(17,5029)(18,33698)(19,70044)
                };
                \label{Involutive-Lattices}

                \addplot[
                color=red,
                mark=square,
                ]
                coordinates {
                	(9,0.12)(10,0.42)(11,0.65)(12,1.36)(13,3.32)(14,8)(15,20)(16,59)(17,128)(18,330)(19,869)(20,2342)(21,6394)(22,18012)(23,51956)
                };
                \label{Tarski-Algebras}

                \addplot[
                color=darkblue,
                mark=otimes,
                ]
                coordinates {
                        (7,1.64)(8,11.07)(9,93)(10,891)(11,8715)(12,100167)
                };
                \label{M-zeroids}
                
				\node [draw,fill=white] at (rel axis cs: 0.72,0.12) {\shortstack[l]{
						\ref{Tarski-Algebras} Tarski Algebras \\
						\ref{Involutive-Lattices} Involutive Lattices \\
						\ref{M-zeroids} M-zeroids }};
        \end{axis}
\end{tikzpicture}
		\label{fig:cbif:speed}
\end{figure}
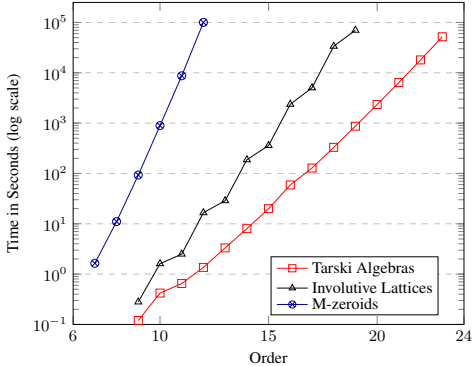

\cbif speeds up the model enumeration process for Tarski Algebras by 3 to 4 orders of magnitude for higher orders. Furthermore, while
the two-step model finding procedure maxes out on the computer capacity at order 12,  \cbif  progresses to order 23 on the same equipment (see Figure~\ref{fig:cbif:speed}). 

\begin{table}[!htb]
	\centering
	\caption{Two-step Model Finding vs.\ \cbif on Tarski Algebras}\label{table:tarski:speedup}
	\begin{tabular}{rrrrr}
		\toprule
		& & \multicolumn{2}{c}{CPU Time (s)} \\ \cmidrule(r){3-4}
		\multicolumn{1}{p{0.6cm}}{\raggedleft Order}  & \multicolumn{1}{p{2.5cm}}{\raggedleft \#Non-isomorphic Models} & \multicolumn{1}{p{1.2cm}}{\raggedleft Two-step}& \multicolumn{1}{p{0.7cm}}{\raggedleft \cbif}& \multicolumn{1}{p{1.2cm}}{\raggedleft Speedup (\#times)}  \\ \cmidrule{1-5}
		9 & 11 & 10 & 0.12 & 80   \\
		10 & 18 & 103 & 0.42 & 245   \\
		11 & 29 & 1,274 & 0.65 & 1,959   \\
		12 & 49 & 17,912 & 1.36 & 13,170  \\ \bottomrule
	\end{tabular}
\end{table}

Involutive Lattices's axioms are more complex than those of Tarski Algebras. 
Nevertheless, \cbif works just as well, giving a speedup of 3 orders of magnitude in higher orders, and finds all models up to order 19 (see Table~\ref{table:involutive:speedup} and Figure~\ref{fig:cbif:speed}).

		 \begin{table}[!htb]
			\centering
			\caption{Two-step Model Finding vs.\ \cbif on Involutive Lattices}
			\begin{tabular}{rrrrr}
				\toprule
				& & \multicolumn{2}{c}{CPU Time (s)} \\ \cmidrule(r){3-4}
				\multicolumn{1}{p{0.6cm}}{\raggedleft Order}  & \multicolumn{1}{p{2.5cm}}{\raggedleft \#Non-isomorphic Models} & \multicolumn{1}{p{1.2cm}}{\raggedleft Two-step}& \multicolumn{1}{p{0.7cm}}{\raggedleft \cbif}& \multicolumn{1}{p{1.2cm}}{\raggedleft Speedup (\#times)}  \\ \cmidrule{1-5}
				9 & 122 & 37 & 0.28 & 132   \\
				10 & 389 & 354 & 1.61 & 220   \\
				11 & 906 & 3,523 & 2.50 & 1,409   \\
				12 & 3,047 & 96,519 & 16.60 & 5,814 \\ \bottomrule
			\end{tabular}
			\label{table:involutive:speedup}
			\end{table}
Finally, we run experiments on the M-zeroids,
which have the most complex axiomatization among the three.
The results as shown in Table~\ref{table:m:zeroids:speedup} and Figure~\ref{fig:speedup}  confirm the pattern: The higher the order, the more pronounced the advantage of \cbif has over the two-step approach.

\begin{table}[!htb]
			\centering
			\caption{Two-step Model Finding vs.\ \cbif on M-zeroids}
			\begin{tabular}{rrrrr}
				\toprule
				& & \multicolumn{2}{c}{CPU Time (s)} \\ \cmidrule(r){3-4}
				\multicolumn{1}{p{0.5cm}}{\raggedleft Order}  & \multicolumn{1}{p{2.1cm}}{\raggedleft \#Non-iso. Models} & \multicolumn{1}{p{1.1cm}}{\raggedleft Two-step}& \multicolumn{1}{p{0.7cm}}{\raggedleft \cbif}& \multicolumn{1}{p{2cm}}{\raggedleft Speedup (\#times)}  \\ \cmidrule{1-5}
				7 & 315 & 35 & 1.64 & 21   \\
				8 & 1,537 & 980 & 11.07	 & 89   \\
				9 & 8,583 & 3,523 & 92.84 & 365   \\ \bottomrule
			\end{tabular}
			\label{table:m:zeroids:speedup}
		\end{table}

While  \cbif is useful for all problems across the board, 
its most valuable property is its ability to retain its effectiveness in working with higher orders of complex structures.  Our tests show that the processing time by \cbif grows much slower than that of the two-step algorithm as the order goes higher.  While the more complex structures have steeper slopes and take longer to generate models, as expected, they all show the same growth pattern (see Figure~\ref{fig:cbif:speed}).

\begin{figure}[!htb]
	\caption{Speedup w/ \cbif over Two-step Algorithm}\label{fig:speedup}
		\centering

\begin{tikzpicture}[scale=0.75, every node/.style={scale=0.9}]
        \begin{axis}[
                xlabel={Order},
                ylabel={Speedup in Run Time (log scale)},
                xmin=6, xmax=13,
                ymin=10, ymax=20000,
                xtick={6,8,10,12},
                ytick={10,100,1000,10000},
                legend pos=north west,
                ymajorgrids=true,
                ymode=log,
                grid style=dashed,
                ]

                \addplot[
                color=black,
                mark=triangle,
                ]
                coordinates {
                        (9,132)(10,220)(11,1409)(12,5814)
                };
                \label{Involutive-Lattices-speed}

                \addplot[
                color=red,
                mark=square,
                ]
                coordinates {
                	(9,80)(10,245)(11,1959)(12,13170)
                };
                \label{Tarski-Algebras-speed}

                \addplot[
                color=darkblue,
                mark=otimes,
                ]
                coordinates {
                        (7,21)(8,89)(9,365)
                };
                \label{M-zeroids-speed}
                
				\node [draw,fill=white] at (rel axis cs: 0.28,0.88) {\shortstack[l]{
						\ref{Tarski-Algebras-speed} Tarski Algebras \\
						\ref{Involutive-Lattices-speed} Involutive Lattices \\
						\ref{M-zeroids-speed} M-zeroids }};
        \end{axis}
\end{tikzpicture}
\end{figure}
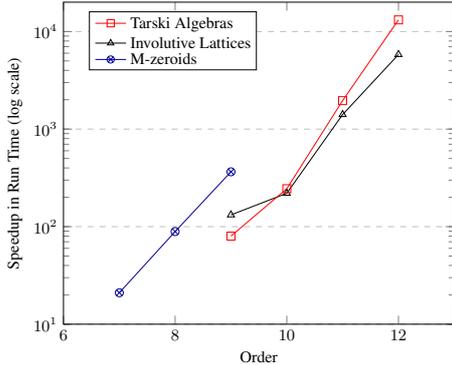

\subsection{New Results}%
\label{sec:NewResults}
 
Here, we present some challenging examples in which \cbif contributes new results to the literature. 
The number of near-rings up to order 13, and the number of Tarski's HSI models up to order 4, have recently been reported in \emph{The On-Line Encyclopedia of Integer Sequences \textregistered} (OEIS)~\cite{oeis} as sequences~\seqnum{A305858} and~\seqnum{A214396}, respectively.  Using the \cbif algorithm, we extend the sequences to order 15 and order 6, respectively (see Table~\ref{table:new:results}). 
These new results are achieved solely with the \cbif algorithm without any knowledge specific to the algebraic structures in question.

Our last experiments give never-reported-before results on two 
algebraic structures derived from loops: C-loops and IP loops.
Phillips \& Vojt{\v e}chovsk{\'y}~\cite{phillips701711c} characterize nonassociative C-loops of orders up to 14, but are not able to enumerate all C-loops of order 16 (there are no nonassociative C-loops of order 15). Slaney \& Ali~\cite{slaney2008generating} think that ``significantly extending the search may raise different challenges for automated reasoning.'' Using \cbif, we successfully enumerate C-loops of order 16 in less than 10 minutes (see Table~\ref{table:new:results}).

For IP loops, Slaney \& Ali~\cite{slaney2008generating} is the first to report successful enumeration of IP loops of order 13 using FINDER~\cite{DBLP:conf/cade/Slaney94a}, which takes a day to crunch out the results. Khan \emph{et al.}~\cite{khan2017enumeration} point out that ``The IP loops having order greater than 13 are not reported in the literature because of the huge search space.'' Using \cbif, we successfully enumerate IP loops of order 14 in less than 2.5 hours, and IP loops of order 15 in 88 hours (see Table~\ref{table:new:results}). 

\begin{table}[!htb]
			\centering
			\caption{New Results Generated by \cbif} 
		\begin{threeparttable}
			\begin{tabular}{lrrr}
				\toprule
				\multicolumn{1}{p{0.6cm}}{Algebra}&
				\multicolumn{1}{p{0.9cm}}{\raggedleft Order}&
				\multicolumn{1}{p{2.9cm}}{\raggedleft \#Non-isomorphic Models}& \multicolumn{1}{p{1.7cm}}{\raggedleft CPU Time (s)} \\ \cmidrule{1-4}
				Near-rings & 14 & 4,537 & 956   \\
				& 15 & 3,817 & 1,524   \\ \cmidrule{1-4}
				Tarski's HSI& 5 & 13,577 & 3   \\
				&6 & 672,740 & 137 \\  \cmidrule{1-4}
				C-loops & 16 & 122 & 572 \\  \cmidrule{1-4}
				IP loops & 14 & 2,104,112 & 8,570 \\
				 & 15 & 40,897,240 & 316,065 \\ \bottomrule
			\end{tabular}
		\end{threeparttable}
		\label{table:new:results}
	\end{table}

The empirical results from \cbif match those in the literature, if available.  Furthermore, they match the results from the 2-step method for lower orders.



\section{Related Work}%
\label{sec:relatedwork}
A practical approach to model finding/enumeration for models of higher orders is the two-step process~\cite{article202206Invariants,araujo_et_al:LIPIcs.CP.2023.8,DBLP:journals/access/Khan20a,prover9-mace4}, where most work is on the elimination of isomorphic models in the second step. McCune~\cite{prover9-mace4} adds \isofilter and \isofiltertwo to the Prover9/Mace4 system to filter out isomorphic models from the outputs of \mace (the first step). \isofilter does brute-force comparisons of models for isomorphism. \isofiltertwo calculates canonical forms of models so that isomorphic models must have the same canonical form.  Neither program is optimized and does not perform adequately except for very small sets of models of small orders. 
Khan~\cite{DBLP:journals/access/Khan20a} presents a 2-step model enumeration algorithm in which the first step uses the CP solver \emph{or-tools}\cite{cpsatlp}, and the second step is then tree-based algorithm (TVMC). In TVMC, each isomorphism class is represented by a branch in the tree.  Each model is checked against all isomorphism classes by traversing this tree only once. They state that ``Our proposed algorithm can enumerate IP loop of order 13 in six hours (21,952 seconds) on an ordinary desktop system.'' In comparison, \cbif takes only 79 seconds to complete the same task on an ordinary computer.  Ara\'{u}jo \emph{et~al.}~\cite{article202206Invariants} observe that some properties of the models are invariant under isomorphisms and devise the invariant-based parallel algorithm to filter out isomorphic models. Nagy and Vojt{\v e}chovsk{\'y}~\cite{loops3.4.1} also implement a  sophisticated quasigroup-specific invariant-based algorithm in the \loops package to handle isomorphism in quasigroups.  This expensive postprocessing step is obviated in \cbif.

Model searching algorithms can generally be classified according to whether they directly search for models, or are based on some transformations of the inputs~\cite{BAUMGARTNER200958}.  Cube-based algorithms are most applicable to the direct-search approach. In this approach, function selection strategy and cell selection strategy have big impacts on the performance of the finite model finders. See for example~\cite{DBLP:journals/jar/AudemardBH06} on the discussion of various function ordering heuristics. \cbif is agnostic to these strategies and hence can be used in conjunction with them. E.g.,~by using suitable function and cell selection strategies, \cbif can generate lexicographically smallest models at no extra costs.

As noted, the first step in the two-step model finding procedure often produces many isomorphic models
that the second step must filter out.  This is particularly true for models defined with FOL because it is an inherent symmetry property of FOL~\cite{DBLP:conf/frocos/RegerR019}. To mitigate this problem, symmetry-breaking algorithms are applied in this step.
There is a bevy of research results on symmetry-breaking~\cite{DBLP:journals/jar/AudemardBH06,Claessen03newtechniques,ijcai/CodishMPS13,articleCrawfordGinsbergLuksRoy1997,szeider_cp21,DBLP:conf/frocos/RegerR019,DBLP:reference/fai/2,DBLP:conf/aaai/Walsh12}.
Many useful partial symmetry-breaking algorithms, such as the LNH and the extended LNH (XLNH)~\cite{Audemard2001XLNH,DBLP:journals/jar/AudemardBH06}, have been widely used. The LNH can be considered as symmetry-breaking with interchangeable values in constraint satisfaction problems (CSP)~\cite{DBLP:conf/ijcai/HentenryckFPA03}. The XLNH is less flexible as it requires the existence of at least one unary operation.  The LNH is implemented in many systems such as \mace, SEM~\cite{DBLP:conf/ijcai/ZhangZ95}, Falcon~\cite{ZhangJ1996Falcon}, and FMSET~\cite{DBLP:journals/fuin/BenhamouH99}.
Algorithms that remove isomorphic cubes, including the \cbif, are compatible with the LNH~\cite{araujo_et_al:LIPIcs.CP.2023.8}.  \cbif, which can be classified as a solution symmetry-breaker (as defined in~\cite{DBLP:journals/constraints/CohenJJPS06}), is a complete symmetry-breaking algorithm and can therefore work without other symmetry-breaking algorithms such as the LNH.  
However, the LNH is a low-overhead predictive algorithm that is particularly effective in removing short isomorphic cubes and hence nicely complements the \cbif.

Another important symmetry-breaking strategy is to prevent the search engine
from the fruitless exploration by adding symmetry-breaking
input clauses~\cite{articleCrawfordGinsbergLuksRoy1997,DBLP:conf/aaai/Walsh12}.
\cbif is agnostic to the additional inputs and hence can be used in conjunction
with them.

The parallel cube algorithm~\cite{araujo_et_al:LIPIcs.CP.2023.8} is yet another effective symmetry-breaking algorithm. Both \cbif and the parallel cube algorithm are based on the fact that isomorphic cubes lead to isomorphic models and hence redundant isomorphic cubes need not be explored.  The parallel cube algorithm is only part of the first step of the two-step algorithm. It only removes isomorphic cubes at some synchronization points, and hence incurs synchronization overheads. Some redundant branches in the search tree may still be searched.  It also requires disk space to hold the intermediate results. Furthermore, unlike \cbif which avails on the power of the \nauty algorithm, it uses less efficient mechanisms such as invariants to detect isomorphic cubes.  It can, however, improve the search speed by doing the searches in parallel, in addition to removing redundant cubes.

Another local symmetry-breaker, DSYM~\cite{DBLP:journals/jar/AudemardBH06}, exploits local symmetries by finding symmetries under invariant partial interpretations (which are invariant cubes). 
DSYM is a predictive algorithm that works at the \emph{parent} level and predicts which of its immediate children will be isomorphic cubes. Consequently, it only detects symmetries under the same subtree. It is thus not an isomorph-free algorithm, although it generates much fewer isomorphic models than using the LNH alone. \cbif, on the other hand, detects both global and local symmetries the same way, and removes all symmetries.
Nevertheless, \cbif is compatible with DSYM, so they can be used together.


Finite model enumeration can be posed as a \emph{constraint programming (CP)}
task~\cite{DBLP:journals/access/Khan20a}. CP solvers abound, e.g., Minion~\cite{DBLP:conf/ecai/GentJM06},
Gecode~\cite{Nielsen06parallelsearch}, and Google's CP-SAT~\cite{cpsatlp}. However, they are usually not isomorph-free model generators.
As pointed out in~\cite{araujo_et_al:LIPIcs.CP.2023.8}, to effectively add
\emph{symmetry-breaking} constraints such as lex-leaders to a CP solver often requires
deep knowledge of the solver, including its input language, \emph{and} the problem at hand.
For example, Distler \emph{et. al.}~\cite{DBLP:conf/cp/DistlerJKK12} posts $2\times 10!$ (over 7 millions) symmetry-breaking constraints in using Minion to find semigroups of order 10.
That is, $\Omega(n!)$ constraints are needed for isomorph-free model generation of models of order $n$. 
The \cbif, on the other hand, requires no additional hand-crafted or program-generated symmetry-breaking constraints.  It can be applied to searching in any structures as long as the finite model finder does direct search. 
%

Some finite model finders, such as 
SEMD~\cite{ZhangJia2006Iso}, also completely suppress isomorphic models in the search process. Like \cbif, SEMD also follows the recorded objects paradigm. In SEMD,  decision sequences for  branches of the search tree are recorded.  The search branch can be discarded if its decision sequence has already appeared in some other search branches.  
It is not clear how well it works with the complex search problems such as those presented in this paper.  They report results up to order 4 for the Tarski's HSI Problem, but we find all models of order 6 for the same problem in just 137 seconds (see Table~\ref{table:new:results}).

\section{Conclusions and Future Work}%
\label{sec:futurework}
In this paper, we introduce the efficient isomorph-free model enumeration algorithm, \cbif, which has wide applications in many research areas such as computational algebra. It enables us to add many new results in computational algebra to the literature (Section~\ref{sec:NewResults}).
It exploits the powers hidden in different representations (operation tables and graphs) of the structures. Not only does it run much faster than the traditional two-step model finders, it also generates no intermediate outputs that take up disk space, which often exceed the limits of the system. This greatly enhances the capability of the finite model searcher in dealing with large search spaces. Moreover, it handles exceptionally well structures with complex axiomatization and retains its power in higher domain sizes.  Last but not least, it is a complete symmetry breaker that  also works with other symmetry-breaking algorithms such as the LNH, and it is easy to be incorporated into any traditional direct-search finite model finder such as \mace.%

Future research will focus on finding the best cell
selection strategy, on fast conversion of models to graphs, and on compact representation of canonical models.






\begin{ack}
The authors would like to thank Prof.~Brendan D.~McKay for advising us how to construct the graphs of models.\\
\noindent
The results were supported by the Ministry of Education, Youth and Sports within the dedicated program ERC~CZ under the project \emph{POSTMAN} no.~LL1902.
\noindent
The research was co-funded by the European Union under the project \emph{ROBOPROX}
(reg.~no.~CZ.02.01.01/00/22\_008/0004590) and
Funda{\c c}\~ao para a Ci\^encia e a Tecnologia,
through the projects UIDB/00297{-}/2020
(CMA), PTDC/MAT-PUR/31174/2017, UIDB/04621/2020 and UIDP/04621/2020.
\end{ack}



\bibliography{refs}

\begin{thebibliography}{42}
\providecommand{\natexlab}[1]{#1}
\providecommand{\url}[1]{\texttt{#1}}
\expandafter\ifx\csname urlstyle\endcsname\relax
  \providecommand{\doi}[1]{doi: #1}\else
  \providecommand{\doi}{doi: \begingroup \urlstyle{rm}\Url}\fi

\bibitem[Ara{\'u}jo et~al.(2022)Ara{\'u}jo, Chow, and
  Janota]{article202206Invariants}
J.~Ara{\'u}jo, C.~Chow, and M.~Janota.
\newblock Boosting isomorphic model filtering with invariants.
\newblock \emph{Constraints}, 27\penalty0 (3):\penalty0 360--379, Jul 2022.
\newblock ISSN 1572-9354.
\newblock \doi{10.1007/s10601-022-09336-x}.

\bibitem[Ara{\'{u}}jo et~al.(2022)Ara{\'{u}}jo, Matos, and Ramires]{marciedb}
J.~Ara{\'{u}}jo, D.~Matos, and J.~Ramires.
\newblock {MarcieDB}: a model and theory database.
\newblock \url{https://marciedb.pythonanywhere.com}, 2022.

\bibitem[Ara\'{u}jo et~al.(2023)Ara\'{u}jo, Chow, and
  Janota]{araujo_et_al:LIPIcs.CP.2023.8}
J.~a. Ara\'{u}jo, C.~Chow, and M.~Janota.
\newblock Symmetries for cube-and-conquer in finite model finding.
\newblock In R.~H.~C. Yap, editor, \emph{29th International Conference on
  Principles and Practice of Constraint Programming (CP 2023)}, volume 280 of
  \emph{Leibniz International Proceedings in Informatics (LIPIcs)}, pages
  8:1--8:19, Dagstuhl, Germany, 2023. Schloss Dagstuhl -- Leibniz-Zentrum
  f{\"u}r Informatik.
\newblock ISBN 978-3-95977-300-3.
\newblock \doi{10.4230/LIPIcs.CP.2023.8}.
\newblock URL \url{https://drops.dagstuhl.de/opus/volltexte/2023/19045}.

\bibitem[Audemard and Henocque(2001)]{Audemard2001XLNH}
G.~Audemard and L.~Henocque.
\newblock The {eXtended} least number heuristic.
\newblock In R.~Gor{\'{e}}, A.~Leitsch, and T.~Nipkow, editors, \emph{Automated
  Reasoning, First International Joint Conference, {IJCAR}}, volume 2083 of
  \emph{Lecture Notes in Computer Science}, pages 427--442, Berlin, Heidelberg,
  2001. Springer.
\newblock \doi{10.1007/3-540-45744-5\_35}.

\bibitem[Audemard et~al.(2006)Audemard, Benhamou, and
  Henocque]{DBLP:journals/jar/AudemardBH06}
G.~Audemard, B.~Benhamou, and L.~Henocque.
\newblock Predicting and detecting symmetries in {FOL} finite model search.
\newblock \emph{J. Autom. Reason.}, 36\penalty0 (3):\penalty0 177--212, 2006.
\newblock \doi{10.1007/s10817-006-9040-3}.

\bibitem[Baumgartner et~al.(2009)Baumgartner, Fuchs, {de Nivelle}, and
  Tinelli]{BAUMGARTNER200958}
P.~Baumgartner, A.~Fuchs, H.~{de Nivelle}, and C.~Tinelli.
\newblock Computing finite models by reduction to function-free clause logic.
\newblock \emph{Journal of Applied Logic}, 7\penalty0 (1):\penalty0 58--74,
  2009.
\newblock ISSN 1570-8683.
\newblock \doi{https://doi.org/10.1016/j.jal.2007.07.005}.
\newblock URL
  \url{https://www.sciencedirect.com/science/article/pii/S1570868307000638}.
\newblock Special Issue: Empirically Successful Computerized Reasoning.

\bibitem[Benhamou and Henocque(1999)]{DBLP:journals/fuin/BenhamouH99}
B.~Benhamou and L.~Henocque.
\newblock A hybrid method for finite model search in equational theories.
\newblock \emph{Fundam. Informaticae}, 39\penalty0 (1-2):\penalty0 21--38,
  1999.
\newblock \doi{10.3233/FI-1999-391202}.

\bibitem[Burris and Lee(1993)]{02661029-c12f-3b75-a049-79273fea242e}
S.~Burris and S.~Lee.
\newblock Tarski's high school identities.
\newblock \emph{The American Mathematical Monthly}, 100\penalty0 (3):\penalty0
  231--236, 1993.
\newblock ISSN 00029890, 19300972.
\newblock URL \url{http://www.jstor.org/stable/2324454}.

\bibitem[Chow(2023)]{mace4c:source}
C.~Chow.
\newblock Mace4c, 2023.
\newblock URL \url{https://github.com/ChoiwahChow/p9m4}.

\bibitem[Chow(2024{\natexlab{a}})]{experiments:source}
C.~Chow.
\newblock Experiments with {Mace4c}, 2024{\natexlab{a}}.
\newblock URL
  \url{https://github.com/ChoiwahChow/public/blob/main/ecai24_cbif.zip}.

\bibitem[Chow(2024{\natexlab{b}})]{isonaut:source}
C.~Chow.
\newblock Isonaut, 2024{\natexlab{b}}.
\newblock URL \url{https://github.com/ChoiwahChow/isonaut}.

\bibitem[Claessen and Sörensson(2003)]{Claessen03newtechniques}
K.~Claessen and N.~Sörensson.
\newblock New techniques that improve {MACE}-style finite model finding.
\newblock In \emph{Proceedings of the CADE-19 Workshop: Model Computation -
  Principles, Algorithms, Applications}, 2003.

\bibitem[Codish et~al.(2013)Codish, Miller, Prosser, and
  Stuckey]{ijcai/CodishMPS13}
M.~Codish, A.~Miller, P.~Prosser, and P.~J. Stuckey.
\newblock Breaking symmetries in graph representation.
\newblock In F.~Rossi, editor, \emph{{IJCAI} 2013, Proceedings of the 23rd
  International Joint Conference on Artificial Intelligence}, pages 510--516.
  {IJCAI/AAAI}, 2013.
\newblock URL
  \url{http://www.aaai.org/ocs/index.php/IJCAI/IJCAI13/paper/view/6480}.

\bibitem[Cohen et~al.(2006)Cohen, Jeavons, Jefferson, Petrie, and
  Smith]{DBLP:journals/constraints/CohenJJPS06}
D.~A. Cohen, P.~Jeavons, C.~Jefferson, K.~E. Petrie, and B.~M. Smith.
\newblock Symmetry definitions for constraint satisfaction problems.
\newblock \emph{Constraints An Int. J.}, 11\penalty0 (2-3):\penalty0 115--137,
  2006.
\newblock \doi{10.1007/s10601-006-8059-8}.

\bibitem[Crawford et~al.(1996)Crawford, Ginsberg, Luks, and
  Roy]{articleCrawfordGinsbergLuksRoy1997}
J.~M. Crawford, M.~L. Ginsberg, E.~M. Luks, and A.~Roy.
\newblock Symmetry-breaking predicates for search problems.
\newblock In L.~C. Aiello, J.~Doyle, and S.~C. Shapiro, editors,
  \emph{Proceedings of the Fifth International Conference on Principles of
  Knowledge Representation and Reasoning (KR)}, pages 148--159. Morgan
  Kaufmann, 1996.

\bibitem[Distler and Mitchell(2019)]{smallsemi0.6.12}
A.~Distler and J.~Mitchell.
\newblock {Smallsemi}, a library of small semigroups in {GAP}, {V}ersion
  0.6.12.
\newblock {https://gap-packages.github.io/smallsemi/}, 2019.
\newblock {G}AP package.

\bibitem[Distler et~al.(2012)Distler, Jefferson, Kelsey, and
  Kotthoff]{DBLP:conf/cp/DistlerJKK12}
A.~Distler, C.~Jefferson, T.~Kelsey, and L.~Kotthoff.
\newblock The semigroups of order 10.
\newblock In M.~Milano, editor, \emph{Principles and Practice of Constraint
  Programming - CP}, volume 7514, pages 883--899. Springer, 2012.
\newblock \doi{10.1007/978-3-642-33558-7\_63}.
\newblock URL \url{https://doi.org/10.1007/978-3-642-33558-7\_63}.

\bibitem[Gent et~al.(2006)Gent, Jefferson, and Miguel]{DBLP:conf/ecai/GentJM06}
I.~P. Gent, C.~Jefferson, and I.~Miguel.
\newblock Minion: {A} fast scalable constraint solver.
\newblock In G.~Brewka, S.~Coradeschi, A.~Perini, and P.~Traverso, editors,
  \emph{{ECAI}, 17th European Conference on Artificial Intelligence, Including
  Prestigious Applications of Intelligent Systems {(PAIS}), Proceedings},
  volume 141 of \emph{Frontiers in Artificial Intelligence and Applications},
  pages 98--102, Amsterdam, Netherlands, 2006. {IOS} Press.
\newblock URL
  \url{http://www.booksonline.iospress.nl/Content/View.aspx?piid=1654}.

\bibitem[Hentenryck et~al.(2003)Hentenryck, Flener, Pearson, and
  {\AA}gren]{DBLP:conf/ijcai/HentenryckFPA03}
P.~V. Hentenryck, P.~Flener, J.~Pearson, and M.~{\AA}gren.
\newblock Tractable symmetry breaking for {CSPs} with interchangeable values.
\newblock In G.~Gottlob and T.~Walsh, editors, \emph{IJCAI-03, Proceedings of
  the Eighteenth International Joint Conference on Artificial Intelligence},
  pages 277--284. Morgan Kaufmann, 2003.
\newblock URL \url{http://ijcai.org/Proceedings/03/Papers/041.pdf}.

\bibitem[Janota et~al.(2024)Janota, Chow, Ara{\'{u}}jo, Codish, and Vojt{\v
  e}chovsk{\'{y}}]{DBLP:conf/aaai/JanotaC0CV24}
M.~Janota, C.~Chow, J.~Ara{\'{u}}jo, M.~Codish, and P.~Vojt{\v e}chovsk{\'{y}}.
\newblock Sat-based techniques for lexicographically smallest finite models.
\newblock In M.~J. Wooldridge, J.~G. Dy, and S.~Natarajan, editors,
  \emph{Thirty-Eighth {AAAI} Conference on Artificial Intelligence, {AAAI}
  2024, Thirty-Sixth Conference on Innovative Applications of Artificial
  Intelligence, {IAAI} 2024, Fourteenth Symposium on Educational Advances in
  Artificial Intelligence, {EAAI} 2014, February 20-27, 2024, Vancouver,
  Canada}, pages 8048--8056. {AAAI} Press, 2024.
\newblock \doi{10.1609/AAAI.V38I8.28643}.
\newblock URL \url{https://doi.org/10.1609/aaai.v38i8.28643}.

\bibitem[Jia and Zhang(2006)]{ZhangJia2006Iso}
X.~Jia and J.~Zhang.
\newblock A powerful technique to eliminate isomorphism in finite model search.
\newblock In U.~Furbach and N.~Shankar, editors, \emph{Automated Reasoning},
  pages 318--331, Berlin, Heidelberg, 2006. Springer Berlin Heidelberg.
\newblock ISBN 978-3-540-37188-5.

\bibitem[Khan(2020)]{DBLP:journals/access/Khan20a}
M.~A. Khan.
\newblock Efficient enumeration of higher order algebraic structures.
\newblock \emph{{IEEE} Access}, 8:\penalty0 41309--41324, 2020.
\newblock \doi{10.1109/ACCESS.2020.2976876}.
\newblock URL \url{https://doi.org/10.1109/ACCESS.2020.2976876}.

\bibitem[Khan et~al.(2017)Khan, Muhammad, Mohammad, and
  Ali]{khan2017enumeration}
M.~A. Khan, S.~Muhammad, N.~Mohammad, and A.~Ali.
\newblock Enumeration of exponent three {IP} loops.
\newblock \emph{Quasigroups \& Related Systems}, 25\penalty0 (1), 2017.

\bibitem[Kirchweger and Szeider(2021)]{szeider_cp21}
M.~Kirchweger and S.~Szeider.
\newblock {SAT} modulo symmetries for graph generation.
\newblock In L.~D. Michel, editor, \emph{27th International Conference on
  Principles and Practice of Constraint Programming, {CP}}, volume 210 of
  \emph{LIPIcs}, pages 34:1--34:16. Schloss Dagstuhl - Leibniz-Zentrum
  f{\"{u}}r Informatik, 2021.
\newblock \doi{10.4230/LIPIcs.CP.2021.34}.
\newblock URL \url{https://doi.org/10.4230/LIPIcs.CP.2021.34}.

\bibitem[McCune(2003)]{Mccune_2003_tech264}
W.~McCune.
\newblock Mace4 reference manual and guide, August 2003.
\newblock URL \url{https://www.cs.unm.edu/~mccune/prover9/mace4.pdf}.

\bibitem[McCune(2005--2010)]{prover9-mace4}
W.~McCune.
\newblock Prover9 and mace4.
\newblock \verb|http://www.cs.unm.edu/~mccune/prover9/|, 2005--2010.

\bibitem[McKay and Piperno(2014)]{mckay2022nauty}
B.~D. McKay and A.~Piperno.
\newblock Practical graph isomorphism, {II}.
\newblock \emph{J. Symb. Comput.}, 60:\penalty0 94--112, 2014.
\newblock \doi{10.1016/j.jsc.2013.09.003}.
\newblock URL \url{https://doi.org/10.1016/j.jsc.2013.09.003}.

\bibitem[Mi{\v c}ek(2015)]{automatedmodel2015}
R.~Mi{\v c}ek.
\newblock Automated model building.
\newblock Master's thesis, Charles University, Czech Republic, 2015.
\newblock http://hdl.handle.net/20.500.11956/81163.

\bibitem[Nagy and Vojt{\v e}chovsk{\'y}(2018)]{loops3.4.1}
G.~Nagy and P.~Vojt{\v e}chovsk{\'y}.
\newblock {LOOPS}, computing with quasigroups and loops in {GAP}, {V}ersion
  3.4.1.
\newblock {https://gap-packages.github.io/loops/}, Nov 2018.
\newblock Refereed GAP package.

\bibitem[Nielsen(2006)]{Nielsen06parallelsearch}
M.~Nielsen.
\newblock Parallel search in gecode.
\newblock \emph{Technical Report, Gecode}, 2006.

\bibitem[Perron and Didier(2023)]{cpsatlp}
L.~Perron and F.~Didier.
\newblock {CP-SAT}, 2023.
\newblock URL \url{https://developers.google.com/optimization/cp/cp_solver/}.

\bibitem[Phillips and Vojtěchovský(2007)]{phillips701711c}
J.~D. Phillips and P.~Vojtěchovský.
\newblock C-loops: An introduction, 2007.

\bibitem[Reger et~al.(2019)Reger, Riener, and Suda]{DBLP:conf/frocos/RegerR019}
G.~Reger, M.~Riener, and M.~Suda.
\newblock Symmetry avoidance in {MACE}-style finite model finding.
\newblock In A.~Herzig and A.~Popescu, editors, \emph{Frontiers of Combining
  Systems FroCoS}, volume 11715, pages 3--21, Switzerland AG, 2019. Springer.
\newblock \doi{10.1007/978-3-030-29007-8\_1}.

\bibitem[Rossi et~al.(2006)Rossi, van Beek, and Walsh]{DBLP:reference/fai/2}
F.~Rossi, P.~van Beek, and T.~Walsh, editors.
\newblock \emph{Handbook of Constraint Programming}, volume~2 of
  \emph{Foundations of Artificial Intelligence}.
\newblock Elsevier, 2006.
\newblock ISBN 978-0-444-52726-4.

\bibitem[Slaney and Ali(2008)]{slaney2008generating}
J.~Slaney and A.~Ali.
\newblock Generating loops with the inverse property.
\newblock \emph{Proc. of ESARM}, pages 55--66, 2008.

\bibitem[Slaney(1994)]{DBLP:conf/cade/Slaney94a}
J.~K. Slaney.
\newblock {FINDER:} finite domain enumerator - system description.
\newblock In A.~Bundy, editor, \emph{Automated Deduction - CADE-12, 12th
  International Conference on Automated Deduction, Nancy, France, June 26 -
  July 1, 1994, Proceedings}, volume 814 of \emph{Lecture Notes in Computer
  Science}, pages 798--801. Springer, 1994.
\newblock \doi{10.1007/3-540-58156-1\_63}.

\bibitem[Sloane and Inc.(2020)]{oeis}
N.~J.~A. Sloane and T.~O.~F. Inc.
\newblock The on-line encyclopedia of integer sequences, 2020.
\newblock URL \url{http://oeis.org/?language=english}.

\bibitem[{The GAP Group}(2019)]{smallgrp}
{The GAP Group}.
\newblock {GAP -- Groups, Algorithms, and Programming, Version 4.10.2}, 2019.
\newblock URL \url{https://www.gap-system.org/Packages/smallgrp.html}.
\newblock Accessed: yyyy-mm-dd.

\bibitem[Walsh(2012)]{DBLP:conf/aaai/Walsh12}
T.~Walsh.
\newblock Symmetry breaking constraints: Recent results.
\newblock In J.~Hoffmann and B.~Selman, editors, \emph{Proceedings of the
  Twenty-Sixth {AAAI} Conference on Artificial Intelligence, July 22-26, 2012,
  Toronto, Ontario, Canada}. {AAAI} Press, 2012.
\newblock URL
  \url{http://www.aaai.org/ocs/index.php/AAAI/AAAI12/paper/view/4974}.

\bibitem[Zhang and Zhang(2013)]{DBLP:conf/birthday/Zhang013}
H.~Zhang and J.~Zhang.
\newblock {MACE4} and {SEM:} {A} comparison of finite model generators.
\newblock In M.~P. Bonacina and M.~E. Stickel, editors, \emph{Automated
  Reasoning and Mathematics - Essays in Memory of William W. McCune}, volume
  7788 of \emph{Lecture Notes in Computer Science}, pages 101--130. Springer,
  2013.
\newblock \doi{10.1007/978-3-642-36675-8\_5}.

\bibitem[Zhang(1996)]{ZhangJ1996Falcon}
J.~Zhang.
\newblock Constructing finite algebras with {FALCON}.
\newblock \emph{Journal of Automated Reasoning}, 17:\penalty0 1--22, 08 1996.
\newblock \doi{10.1007/BF00247667}.

\bibitem[Zhang and Zhang(1995)]{DBLP:conf/ijcai/ZhangZ95}
J.~Zhang and H.~Zhang.
\newblock {SEM:} a system for enumerating models.
\newblock In \emph{IJCAI}, pages 298--303, 1995.
\newblock URL \url{http://ijcai.org/Proceedings/95-1/Papers/039.pdf}.

\end{thebibliography}

\end{document}